\documentstyle[prl,preprint,aps]{revtex}
\draft
\begin{document}

\title{Localized Excitons and Breaking of Chemical Bonds at III-V
(110) Surfaces}

\author{Oleg Pankratov$^{(*)}$ and Matthias Scheffler}

\address{
Fritz-Haber-Institut der Max-Planck-Gesellschaft,\\
Faradayweg 4-6, D-14195 Berlin-Dahlem, Germany}

\date{Received \today}
\maketitle
\begin{abstract}

Electron-hole excitations in the surface bands of GaAs\,(110) are analyzed
using constrained density-functional theory calculations. The results show
that Frenkel-type autolocalized excitons are formed. The excitons induce a
local surface unrelaxation which results in a strong exciton-exciton
attraction and makes complexes of
two or three electron-hole pairs more favorable than separate excitons.
In such microscopic exciton ``droplets'' the electron density is mainly
concentrated in the dangling orbital of a surface Ga atom whereas the holes
are distributed over the bonds of this atom to its As neighbors thus
weakening the bonding to the substrate. This finding suggests the microscopic
mechanism of a laser-induced emission of neutral Ga atoms from
GaAs and GaP (110) surfaces.\\[0.5cm]

\end{abstract}


\newpage
Surface electron states, in particular for covalently bound
$sp$-electron semiconductors, are very sensitive to the surface geometry.
This may actuate a strong electron-lattice coupling at surfaces.
Coulomb correlations will typically also be enhanced in surface bands due to
weaker screening, reduced dimensionality, and larger effective masses of the
charge carriers.

Recent experiments~\cite{kanasaki,hattori} show that electron-hole ($e$-$h$)
excitations at the clean GaAs\,(110) and GaP\,(110) surfaces can cause
emission of
neutral Ga atoms which indicates a strong interaction of the surface $e$-$h$
pairs with the lattice. In the present paper we investigate an electron density
perturbation and related lattice distortion initiated by ``pumping'' $e$-$h$
excitations into the surface states.

The (110) surface of III-V compounds maintains the $(1 \times 1)$
rectangular symmetry of a bulk (110) crystallographic plane. However it
relaxes as such that the surface cation atoms shift inwards and the anion atoms
shift outwards leaving the cation-anion distance almost unchanged.
The relaxation can be described as quasi-rigid rotation of
As-Ga bonds or as a frozen in ``rotational'' surface phonon.
As concluded from inelastic He scattering~\cite{toennis}, this phonon
has practically no dispersion, which means that the As-Ga pairs in different
surface cells move almost independently.

Density-functional theory calculations reproduce the surface geometry for all
III-V compounds quite accurately and
reveal two surface bands around the fundamental gap - one occupied and one
empty (see Ref.~\cite{alves} and references therein).
The occupied state is
predominantly composed of the anion dangling orbitals and the unoccupied one
consists of the cation dangling orbitals. Since in the following we consider
the GaAs\,(110) surface, we will refer to these bands as to As-like or
Ga-like states. Figure~1 shows the surface band structure of GaAs\,(110),
which we calculated using density functional theory in the local-density
approximation (DFT-LDA). We employ a plane wave basis set with a cutoff of 8
Ry and a slab geometry with a slab thickness of 7 layers and vacuum region
equivalent to 5 layers (further details of the calculational method are given
in Refs.~\cite{alves,stumpf} and~\cite{pankratov4}). Since the single-particle
DFT-LDA eigenvalues, strictly speaking, have no direct physical meaning, to
calculate a measurable band structure one has to replace the
exchange-correlation potential, $V^{\rm xc}$ by the non-local, energy
dependent, complex self-energy, $\Sigma (E,{\bf r,r'})$. However, for GaAs and
its (110) surface it has been shown~\cite{godby,louie,zhu} that in the region
of the fundamental energy gap the quasiparticle correction to LDA eigenvalues
is almost ${\bf k}$-independent, i.e., it shifts the bands almost rigidly.
For conduction bands the shift is about 0.7 eV upwards. For valence bands it
is downwards and an order of magnitude smaller. Our DFT-LDA band
structure
shown in Fig.~1 has been corrected by the self-energy effect on the lower
conduction band states.

The occupied As band agrees well with photoemission data
(e.g.~\cite{huijser}) and the empty Ga band is in a qualitative accord with
inverse photoemission~\cite{straub}. Two-step
photoemission~\cite{haight} and optical reflectivity~\cite{selci-berkovits}
give somewhat different energies of the Ga
state. This might reflect the difference of the physical
processes involved in the measurement, since inverse
photoemission gives the quasiparticle energy
of the {\em empty} state whereas {\em preoccupation} of the Ga state in
two-step photoemission or {\em creation of} $e$-$h$ {\em pairs}
in optical reflectivity measurements implies a presence
of polaron and exciton effects. However, importance of these effects becomes
more apparent in optical experiments
at higher excitation level~\cite{kanasaki,hattori}, which show that laser
irradiation of GaAs and GaP\,(110), though still below the ablation threshold,
causes desorption of Ga atoms. The effect is maximal for the light
polarization in $[1 \bar{1} 1]$ direction, i.e., along the surface Ga-As zigzag
chains and practically vanishes for photons polarized in perpendicular
direction.
As the bulk GaAs lattice has a cubic symmetry, this means that the Ga emission
is related to electron-hole excitations at the surface. The yield-fluence
relation is superlinear, i.e., several $e$-$h$ excitations are needed to
release one atom. To
cause the bond breaking these excitations should first become localized, which
was attributed~\cite{kanasaki,hattori} to surface defects. Our calculations
show that even on a defect-free surface a self-trapping of single, double, and
triple excitons occurs.

The clue to understanding the exciton self-trapping is in the link
between the surface relaxation and the energy position of the surface
states. As it can be seen in Fig.~1, the relaxation pushes the surface states
out of the band gap. It is apparently driven by the energy gain due
to lowering of the occupied As
band. The shift of this band agrees with the relaxation energy
of 0.7 eV, as it results from the total energy calculations. The $e$-$h$ pair
with an electron in the Ga band and a hole in the As band tend
to reduce locally the surface relaxation since this would
shift the electron level downwards and the the hole level
upwards and thus would lower the energy of the excitation. The
both particles become then self-trapped in a potential well created by the
local surface
unrelaxation. The localization region is likely to be very compact - about the
size of the surface elementary cell, since this ``saves'' relaxation energy
and is not opposed by the weak elastic coupling to neighboring cells.
This mechanism of exciton self-trapping is somewhat similar to electron
localization due to the surface polaron effect~\cite{pankratov2} which has
been suggested to be responsible for
non-metallic behavior of the GaAs\,(110) surface covered with alkali-metal
submonolayer~\cite{plummer,whitman}.

A simple effective mass estimate shows that the exciton binding energy is
strongly enhanced in the surface states. The ``surface''
effective Rydberg is
\begin{equation}
Ry_{s}^{*}=4\times {\frac{m_{s}^{*}e^{4}}{2\hbar^2\varepsilon_{s}^{2}}}
\approx 16{\frac{m_{s}^{*}}{m^{*}}}Ry^{*}\quad,
\end{equation}
where the factor 4 is due to the two-dimensional character of the
electron motion and $Ry^{*}= m^{*}e^{4}/2\hbar^2 \varepsilon_{0}^{2} \approx 4$
meV is
a ``bulk'' Rydberg with the bulk dielectric constant $\varepsilon_{0}$.
The
``surface dielectric constant'', $\varepsilon_{s}=(1+\varepsilon_{0})/2$,
inhances the exciton binding energy by another factor
$(\varepsilon_{0}/\varepsilon_{s})^{2}\approx 4$. The coefficient
$m_{s}^{*}/m^{*}$ accounts for the different effective mass at the surface. A
rough estimate for GaAs\,(110) gives $m_{s}^{*}/m^{*}\approx 4$, so that
$Ry_{s}^{*} \approx 0.26$ eV and the corresponding Bohr radius, $a_s^{*}
\approx 8$ {\AA} is about the size of the surface elementary cell. Therefore
the surface
excitons are rather of a Frenkel-type than of a Wannier-Mott type, and the
effective mass approximation is actually inadequate.

We use DFT-LDA to study the $e$-$h$ excitations and their interaction with the
lattice. At first glance such approach appears to be not justified,
since DFT is a ground state theory. However, by introducing
appropriate constraints (see e.g. \cite{dederichs}) it is possible to get a
meaningful description of the localized $e$-$h$ excitations in the DFT-LDA
formalism.

In the Kohn-Sham scheme of DFT the total electron density $n({\bf r})$ is
constructed from the auxiliary one-particle wave functions $\phi_{i}({\bf
r})$:
\begin{equation}
n({\bf r})=\sum_{i}n_{i}|\phi_{i}({\bf r})|^{2} \quad.
\end{equation}
The occupation numbers  $n_{i}$ are equal to two for states below the
Fermi energy and to zero above (we presume that each state is spin-degenerate).
Although the Kohn-Sham orbitals $\phi_{i}({\bf r})$ do not have
a direct physical meaning, for GaAs and its (110) surface they are in fact
very close to the actual quasiparticle wave functions in the energy range
around
the band gap~\cite{godby,louie,zhu}. We therefore presume, that setting the
occupation number of one level in a valence surface state to $n_{i}=1$
and occupying one level in the empty surface Ga state we adequately
represent the electron density of a $e$-$h$ excitation. This means that the
density should be determined via a self-consistent solution of the DFT-LDA
equations for $\phi_{i}({\bf r})$ using Eq.~2 with the just noted
constraint on the occupation numbers.

Once the exciton is formed, the electron and hole wave functions do not
separately obey translational invariance but only the whole exciton
wave function does. Therefore if the Kohn-Sham orbitals in Eq.~2 are
required to be Bloch states, the exciton correlations will be missing.
We perform calculations using a ($2 \times 2$) surface cell to allow for
the umklapp processes with wave vectors $\overline{\Gamma}\;\overline{\rm X}$
and $\overline{\Gamma}\;\overline{\rm X'}$ (see upper panel in Fig.~1). These
processes via
the $e$-$h$ Coulomb interaction should help building out of the eigenfunctions
of the ($1 \times 1$) periodical surface the bound $e$-$h$ state. Let us first
consider
the $e$-$h$ pair in point $\overline{\rm X}$, which determines the absorption
edge for optical transitions between
the surface states~\cite{louie}. Figure 2~(a) shows the change of a total
electron density due to this excitation.
No tendency to formation of the
$e$-$h$ bound state is seen and the excitation energy $E_{\rm X}$ is only
slightly
(by about 0.08 eV) smaller than the band gap. This result is actually
apparent, since in our ($2 \times 2$) supercell the $\overline{\rm X}$-states
are allowed to mix only with the states in
$\overline{\Gamma},\overline{\rm X'}$, and $\overline{\rm M}$. In all these
points the Ga-band and As-band wave functions are equally distributed
between, respectively, the Ga and As dangling orbitals and cannot form a
tightly
bound state. In contrast, the $e$-$h$ pair in $\overline{\rm L}$-point leads
to a
localized charge perturbation, as shown in Fig.~2~(b). The reason is that the
eigenfunctions in the four $\overline{\rm L}$-points, connected by the ``new''
reciprocal lattice vectors, can be combined in a wave function with nodes at
all Ga (or, respectively, As) sites in a ($2 \times 2$) cell except
one~\cite{pankratov2,pankratov3}. It is clearly seen in Fig.~2~(b) that
electron and hole form a compact surface exciton. The localized charge
distribution arizes automatically during iterations to self-consistency,
although the ``starting'' charge distribution can look very similar to the one
shown in Fig.~2~(a). The exciton density in Fig.~2~(b) was obtained
maintaining the reflection symmetry in the plane through the Ga atom.
If, instead, a reflection symmetry for a plane through the As atom is
requested, the hole becomes entirely
localized and electron is smeared over surrounding Ga orbitals. This case,
however, corresponds to higher excitation energy, since the exciton-induced
lattice unrelaxation is much smaller (see below). The total energy calculation
shows that the $\overline{\rm L}$-excitation costs 3.1 eV. Comparison with the
energy gap in $\overline{\rm L}$-point (3.6 eV) gives the exciton binding
energy  about 0.5 eV.

Figures~3 (a-d) show the side view at the localized
$\overline{\rm L}$-excitation in the plane indicated by the dashed-dotted line
in
Fig.~2 (b). A localization of both an electron (solid contour lines) and a
hole (dashed contour lines) is apparent. In Fig.~3 (a) the ideal relaxed
surface geometry is kept fixed, i.e. it shows the side view at
the charge distribution of the Fig.~2 (b). In the Figs.~3 (b-d) the
lattice has been relaxed according to the excited electron density
distribution. Therewith we assume that the exciton lifetime is longer than the
inverse surface phonon frequency, $\approx 0.4$ ps. For a single $e$-$h$ pair
[Fig.~3 (b)] the
lattice relaxation lowers the excitation energy down to 2.7 eV. In case of two
and three excited $e$-$h$ pairs [Fig.~3 (c,d)] we find that energetically
optimal density distribution corresponds to double or triple exciton complexes
localized around the same Ga site, but not to separate independent
excitons.

Figure~4 illustrates that this strong exciton-exciton attraction
is almost entirely due to the lattice distortion.
If excitons were not interacting, the energy of two or three
$e$-$h$ pairs would be, respectively, two or three times larger than that of a
single exciton. As the upper curves in Fig.~4 (for both $\overline{\rm X}$ and
$\overline{\rm L}$-excitations) show, this simple proportionality is slightly
violated due to the Coulomb interaction. The lower curves demonstrate a
dramatic effect of the lattice relaxation on $\overline{\rm L}$-exciton and a
very weak one
for $\overline{\rm X}$-excitation. Whereas a single $\overline{\rm L}$-exciton
is energetically more costly than a delocalized $\overline{\rm X}$-excitation,
in the case of two or three excited $e$-$h$ pairs the localized double or
triple exciton complexes become lower in energy. Therefore
at sufficiently high excitation level (which is actually
determined by the exciton kinetics) we expect that the homogeneous state
becomes unstable, and autolocalized double and triple
$e$-$h$ complexes are formed. From Fig.~3(b-d) a large displacement
of the surface Ga atom which holds extra electrons is evident.
The electron density redistribution suggests a weakening of the bond
of this atom to the substrate as a consequence
of penetration of the hole beneath it. It may be then possible that this
surface atom will be set free and emitted as neutral Ga atom when excitons
recombine and excitation energy transforms into the local surface vibrations.

We gratefully acknowledge discussions with R.~Car, J.~Lapeyre and
E.W.~Plummer. This work was supported in part by the Volkswagen-Stiftung and by
Sfb 296 of the DFG.\\

\newpage
\begin{figure}
\caption{Calculated energy band structure of GaAs\,(110) surface along the
perimeter and inside the surface Brillouin zone (SBZ). The shaded regions are
the projected bulk bands. Dashed lines show surface bands obtained for
unrelaxed surface. The energy zero is set at the top of the bulk valence band.
The upper panel displays the top view at the GaAs\,(110)
surface and the SBZ. The broken lines enclose the smaller SBZ of the $(2
\times 2)$ surface cell.}
\label{fig1}
\end{figure}

\begin{figure}
\caption{Top view at the exciton-induced change of the electron density:
(a) an $\overline{\rm X}$-exciton and (b) an $\overline{\rm L}$-exciton. We
display a cut through the (110) plane 0.6 {\AA} above the surface, which runs
through the
middle of the Ga dangling orbitals. Small circles stand for Ga
atoms, large - for As atoms. The lattice is kept at the ideal ground-state
configuration. An increase of the density is
shown by solid lines (electron), a decrease by dashed lines (hole). Note the
difference in scale:
for (a) the density step is $0.5\times10^{-3}{\rm bohr^{-3}}$, for (b) it
is $1\times10^{-3}{\rm bohr^{-3}}$. The dashed-dotted line defines the
orientation of the orthogonal plane used for the side view in Fig.~3}
\label{fig2}
\end{figure}

\begin{figure}
\caption{Side view at the exciton-induced electron density for $\overline{\rm
L}$-excitations. Small circles are Ga atoms, large - As atoms.
(a) A single excited $e$-$h$ pair. The lattice is kept frozen at the clean
surface
ground-state geometry; (b) Same as in (a) but for a relaxed lattice; (c) Double
$e$-$h$  excitation in relaxed lattice; (d) Triple $e$-$h$  excitation in
relaxed lattice. Note the difference in scale:
for (a) the density step is $1\times10^{-3}{\rm bohr^{-3}}$, for (b)-(d) it is
$4\times10^{-3}{\rm bohr^{-3}}$.}
\label{fig3}
\end{figure}

\begin{figure}
\caption{
Excitation energies for different numbers of $e$-$h$ pairs in a $2 \times 2$
surface supercell. $\overline{\rm X}$-excitations (dashed lines) induce a
delocalized electron-hole density; $\overline{\rm L}$-excitations are
localized:
these are a
single $\overline{\rm L}$-exciton and double and triple $e$-$h$ complexes
around
the same Ga site. The upper curves for both $\overline{\rm X}$-  and
$\overline{\rm L}$-excitations have been calculated with frozen lattice,
the lower curves -- with relaxed lattice.}
\label{fig4}
\end{figure}
\end{document}